\documentclass[twocolumn,letterpaper,aps,prc,longbibliography,superscriptaddress,nofootinbib,floatfix]{revtex4-2}

\usepackage{graphicx}   
\usepackage{xspace}     
\usepackage{amsmath}     

\begin{document}

\title{Measurement of elliptic flow of J$/\psi$ in $\sqrt{s_{_{NN}}}=200$ 
GeV Au$+$Au collisions \\
at forward rapidity}

\newcommand{\abilene}{Abilene Christian University, Abilene, Texas 79699, USA}
\newcommand{\augie}{Department of Physics, Augustana University, Sioux Falls, South Dakota 57197, USA}
\newcommand{\banaras}{Department of Physics, Banaras Hindu University, Varanasi 221005, India}
\newcommand{\barc}{Bhabha Atomic Research Centre, Bombay 400 085, India}
\newcommand{\baruch}{Baruch College, City University of New York, New York, New York, 10010 USA}
\newcommand{\bnlcoll}{Collider-Accelerator Department, Brookhaven National Laboratory, Upton, New York 11973-5000, USA}
\newcommand{\bnlphys}{Physics Department, Brookhaven National Laboratory, Upton, New York 11973-5000, USA}
\newcommand{\caucr}{University of California-Riverside, Riverside, California 92521, USA}
\newcommand{\charlesczech}{Charles University, Faculty of Mathematics and Physics, 180 00 Troja, Prague, Czech Republic}
\newcommand{\ciae}{Science and Technology on Nuclear Data Laboratory, China Institute of Atomic Energy, Beijing 102413, People's Republic of China}
\newcommand{\cns}{Center for Nuclear Study, Graduate School of Science, University of Tokyo, 7-3-1 Hongo, Bunkyo, Tokyo 113-0033, Japan}
\newcommand{\colorado}{University of Colorado, Boulder, Colorado 80309, USA}
\newcommand{\columbia}{Columbia University, New York, New York 10027 and Nevis Laboratories, Irvington, New York 10533, USA}
\newcommand{\czechtech}{Czech Technical University, Zikova 4, 166 36 Prague 6, Czech Republic}
\newcommand{\debrecen}{Debrecen University, H-4010 Debrecen, Egyetem t{\'e}r 1, Hungary}
\newcommand{\elte}{ELTE, E{\"o}tv{\"o}s Lor{\'a}nd University, H-1117 Budapest, P{\'a}zm{\'a}ny P.~s.~1/A, Hungary}
\newcommand{\ewha}{Ewha Womans University, Seoul 120-750, Korea}
\newcommand{\fsu}{Florida State University, Tallahassee, Florida 32306, USA}
\newcommand{\gsu}{Georgia State University, Atlanta, Georgia 30303, USA}
\newcommand{\hiroshima}{Physics Program and International Institute for Sustainability with Knotted Chiral Meta Matter (SKCM2), Hiroshima University, Higashi-Hiroshima, Hiroshima 739-8526, Japan}
\newcommand{\howard}{Department of Physics and Astronomy, Howard University, Washington, DC 20059, USA}
\newcommand{\hunrenatomki}{HUN-REN ATOMKI, H-4026 Debrecen, Bem t{\'e}r 18/c, Hungary}
\newcommand{\ihepprot}{IHEP Protvino, State Research Center of Russian Federation, Institute for High Energy Physics, Protvino, 142281, Russia}
\newcommand{\illuiuc}{University of Illinois at Urbana-Champaign, Urbana, Illinois 61801, USA}
\newcommand{\inrras}{Institute for Nuclear Research of the Russian Academy of Sciences, prospekt 60-letiya Oktyabrya 7a, Moscow 117312, Russia}
\newcommand{\instpasczech}{Institute of Physics, Academy of Sciences of the Czech Republic, Na Slovance 2, 182 21 Prague 8, Czech Republic}
\newcommand{\isu}{Iowa State University, Ames, Iowa 50011, USA}
\newcommand{\jaea}{Advanced Science Research Center, Japan Atomic Energy Agency, 2-4 Shirakata Shirane, Tokai-mura, Naka-gun, Ibaraki-ken 319-1195, Japan}
\newcommand{\jeonbuk}{Jeonbuk National University, Jeonju, 54896, Korea}
\newcommand{\jyvaskyla}{Helsinki Institute of Physics and University of Jyv{\"a}skyl{\"a}, P.O.Box 35, FI-40014 Jyv{\"a}skyl{\"a}, Finland}
\newcommand{\kek}{KEK, High Energy Accelerator Research Organization, Tsukuba, Ibaraki 305-0801, Japan}
\newcommand{\korea}{Korea University, Seoul 02841, Korea}
\newcommand{\kurchatov}{National Research Center ``Kurchatov Institute", Moscow, 123098 Russia}
\newcommand{\kyoto}{Kyoto University, Kyoto 606-8502, Japan}
\newcommand{\lawllnl}{Lawrence Livermore National Laboratory, Livermore, California 94550, USA}
\newcommand{\losalamos}{Los Alamos National Laboratory, Los Alamos, New Mexico 87545, USA}
\newcommand{\lund}{Department of Physics, Lund University, Box 118, SE-221 00 Lund, Sweden}
\newcommand{\lyon}{IPNL, CNRS/IN2P3, Univ Lyon, Universit{\'e} Lyon 1, F-69622, Villeurbanne, France}
\newcommand{\maryland}{University of Maryland, College Park, Maryland 20742, USA}
\newcommand{\mass}{Department of Physics, University of Massachusetts, Amherst, Massachusetts 01003-9337, USA}
\newcommand{\mate}{MATE, Laboratory of Femtoscopy, K\'aroly R\'obert Campus, H-3200 Gy\"ongy\"os, M\'atrai \'ut 36, Hungary}
\newcommand{\michigan}{Department of Physics, University of Michigan, Ann Arbor, Michigan 48109-1040, USA}
\newcommand{\miss}{Mississippi State University, Mississippi State, Mississippi 39762, USA}
\newcommand{\muhlenberg}{Muhlenberg College, Allentown, Pennsylvania 18104-5586, USA}
\newcommand{\nara}{Nara Women's University, Kita-uoya Nishi-machi Nara 630-8506, Japan}
\newcommand{\natmephi}{National Research Nuclear University, MEPhI, Moscow Engineering Physics Institute, Moscow, 115409, Russia}
\newcommand{\newmex}{University of New Mexico, Albuquerque, New Mexico 87131, USA}
\newcommand{\nmsu}{New Mexico State University, Las Cruces, New Mexico 88003, USA}
\newcommand{\northcg}{Physics and Astronomy Department, University of North Carolina at Greensboro, Greensboro, North Carolina 27412, USA}
\newcommand{\ohio}{Department of Physics and Astronomy, Ohio University, Athens, Ohio 45701, USA}
\newcommand{\ornl}{Oak Ridge National Laboratory, Oak Ridge, Tennessee 37831, USA}
\newcommand{\orsay}{IPN-Orsay, Univ.~Paris-Sud, CNRS/IN2P3, Universit\'e Paris-Saclay, BP1, F-91406, Orsay, France}
\newcommand{\peking}{Peking University, Beijing 100871, People's Republic of China}
\newcommand{\pnpi}{PNPI, Petersburg Nuclear Physics Institute, Gatchina, Leningrad region, 188300, Russia}
\newcommand{\pusan}{Pusan National University, Pusan 46241, Korea}
\newcommand{\riken}{RIKEN Nishina Center for Accelerator-Based Science, Wako, Saitama 351-0198, Japan}
\newcommand{\rikjrbrc}{RIKEN BNL Research Center, Brookhaven National Laboratory, Upton, New York 11973-5000, USA}
\newcommand{\rikkyo}{Physics Department, Rikkyo University, 3-34-1 Nishi-Ikebukuro, Toshima, Tokyo 171-8501, Japan}
\newcommand{\saispbstu}{Saint Petersburg State Polytechnic University, St.~Petersburg, 195251 Russia}
\newcommand{\seoulnat}{Department of Physics and Astronomy, Seoul National University, Seoul 151-742, Korea}
\newcommand{\stonybrkc}{Chemistry Department, Stony Brook University, SUNY, Stony Brook, New York 11794-3400, USA}
\newcommand{\stonycrkp}{Department of Physics and Astronomy, Stony Brook University, SUNY, Stony Brook, New York 11794-3800, USA}
\newcommand{\tenn}{University of Tennessee, Knoxville, Tennessee 37996, USA}
\newcommand{\titech}{Department of Physics, Tokyo Institute of Technology, Oh-okayama, Meguro, Tokyo 152-8551, Japan}
\newcommand{\tsukuba}{Tomonaga Center for the History of the Universe, University of Tsukuba, Tsukuba, Ibaraki 305, Japan}
\newcommand{\usmma}{United States Merchant Marine Academy, Kings Point, New York 11024, USA}
\newcommand{\vandy}{Vanderbilt University, Nashville, Tennessee 37235, USA}
\newcommand{\weizmann}{Weizmann Institute, Rehovot 76100, Israel}
\newcommand{\wigner}{Institute for Particle and Nuclear Physics, HUN-REN Wigner Research Centre for Physics, (HUN-REN Wigner RCP, RMI), H-1525 Budapest 114, POBox 49, Budapest, Hungary}
\newcommand{\yonsei}{Yonsei University, IPAP, Seoul 120-749, Korea}
\newcommand{\zagreb}{Department of Physics, Faculty of Science, University of Zagreb, Bijeni\v{c}ka c.~32 HR-10002 Zagreb, Croatia}
\newcommand{\zambia}{Department of Physics, School of Natural Sciences, University of Zambia, Great East Road Campus, Box 32379, Lusaka, Zambia}
\affiliation{\abilene}
\affiliation{\augie}
\affiliation{\banaras}
\affiliation{\barc}
\affiliation{\baruch}
\affiliation{\bnlcoll}
\affiliation{\bnlphys}
\affiliation{\caucr}
\affiliation{\charlesczech}
\affiliation{\ciae}
\affiliation{\cns}
\affiliation{\colorado}
\affiliation{\columbia}
\affiliation{\czechtech}
\affiliation{\debrecen}
\affiliation{\elte}
\affiliation{\ewha}
\affiliation{\fsu}
\affiliation{\gsu}
\affiliation{\hiroshima}
\affiliation{\howard}
\affiliation{\hunrenatomki}
\affiliation{\ihepprot}
\affiliation{\illuiuc}
\affiliation{\inrras}
\affiliation{\instpasczech}
\affiliation{\isu}
\affiliation{\jaea}
\affiliation{\jeonbuk}
\affiliation{\jyvaskyla}
\affiliation{\kek}
\affiliation{\korea}
\affiliation{\kurchatov}
\affiliation{\kyoto}
\affiliation{\lawllnl}
\affiliation{\losalamos}
\affiliation{\lund}
\affiliation{\lyon}
\affiliation{\maryland}
\affiliation{\mass}
\affiliation{\mate}
\affiliation{\michigan}
\affiliation{\miss}
\affiliation{\muhlenberg}
\affiliation{\nara}
\affiliation{\natmephi}
\affiliation{\newmex}
\affiliation{\nmsu}
\affiliation{\northcg}
\affiliation{\ohio}
\affiliation{\ornl}
\affiliation{\orsay}
\affiliation{\peking}
\affiliation{\pusan} 
\affiliation{\pnpi}
\affiliation{\riken}
\affiliation{\rikjrbrc}
\affiliation{\rikkyo}
\affiliation{\saispbstu}
\affiliation{\seoulnat}
\affiliation{\stonybrkc}
\affiliation{\stonycrkp}
\affiliation{\tenn}
\affiliation{\titech}
\affiliation{\tsukuba}
\affiliation{\usmma}
\affiliation{\vandy}
\affiliation{\weizmann}
\affiliation{\wigner}
\affiliation{\yonsei}
\affiliation{\zagreb}
\affiliation{\zambia}
\author{N.J.~Abdulameer} \affiliation{\debrecen} \affiliation{\hunrenatomki}
\author{U.~Acharya} \affiliation{\gsu}
\author{A.~Adare} \affiliation{\colorado} 
\author{C.~Aidala} \affiliation{\michigan} 
\author{N.N.~Ajitanand} \altaffiliation{Deceased} \affiliation{\stonybrkc} 
\author{Y.~Akiba} \email[PHENIX Spokesperson: ]{akiba@rcf.rhic.bnl.gov} \affiliation{\riken} \affiliation{\rikjrbrc}
\author{M.~Alfred} \affiliation{\howard} 
\author{S.~Antsupov} \affiliation{\saispbstu}
\author{K.~Aoki} \affiliation{\kek} \affiliation{\riken} 
\author{N.~Apadula} \affiliation{\isu} \affiliation{\stonycrkp} 
% ...
\collaboration{PHENIX Collaboration}  \noaffiliation

\date{\today}


\begin{abstract}


We report the first measurement of the azimuthal anisotropy of J$/\psi$ 
at forward rapidity ($1.2<|\eta|<2.2$) in Au$+$Au collisions at 
$\sqrt{s_{_{NN}}}=200$ GeV at the Relativistic Heavy Ion Collider. The 
data were collected by the PHENIX experiment in 2014 and 2016 with 
integrated luminosity of 14.5~nb$^{-1}$. The second Fourier coefficient 
($v_2$) of the azimuthal distribution of $J/\psi$ is determined as a 
function of the transverse momentum ($p_T$) using the event-plane 
method. The measurements were performed for several selections of 
collision centrality: 0\%--50\%, 10\%--60\%, and 10\%-40\%. We find 
that in all cases the values of $v_2(p_T)$, which quantify the elliptic 
flow of J$/\psi$, are consistent with zero. The results are consistent 
with measurements at midrapidity, indicating no significant elliptic 
flow of the J$/\psi$ within the quark-gluon-plasma medium at collision 
energies of $\sqrt{s_{_{NN}}}=200$ GeV.

\end{abstract}
\maketitle

\section{Introduction}
\label{sec:intro}

A state of deconfined quarks and gluons, the quark-gluon plasma (QGP), 
is formed in heavy ion collisions at the BNL Relativistic Heavy Ion 
Collider 
(RHIC)~\cite{PHENIX:2004vcz,BRAHMS:2004adc,PHOBOS:2004zne,STAR:2005gfr}. 
The QGP has been found to exhibit a near-perfect-fluid behavior 
inferred from the strong correlations between the particles produced in 
the collisions~\cite{Heinz:2008tv}. These correlations are observed as 
an anisotropy in the final-state particle azimuthal distributions and 
are quantified by the components of the Fourier expansion of the 
invariant particle yields~\cite{Poskanzer:1998yz}.

\begin{small}
\begin{equation}
 E\frac{d^3N}{d^3p}=\frac{1}{2\pi}\frac{d^2N}{p_Tdp_Tdy}\left[1+\sum^\infty_{n=1}2v_n\cos[n(\phi-\Psi_n)]\right],
\end{equation}
\end{small}

\noindent where $\Psi_n$ represents the $n^{th}$-order symmetry plane 
angle, and $v_n$ represent the flow coefficient for the 
$n^{th}$-harmonic. The second harmonic, $v_2$, known as ``elliptic 
flow" is associated with the initial geometry of the overlap region of 
the colliding nuclei, which drives the anisotropy of the final-state 
particle distribution through anisotropic pressure gradients in the 
QGP.

Measuring the elliptic flow of heavy-flavor particles is of particular 
interest because due to their relatively large masses, the charm and 
beauty quarks may be thermodynamically distinct from the QGP medium. 
This feature, in addition to their relatively long lifetimes and 
production in hard-scattering events during the initial stages of the 
collision, make them excellent probes for testing the properties of the 
QGP medium.

Previous measurements show a significant $v_2$ of charmed hadrons in 
Au$+$Au collisions~\cite{Hachiya:2019stg,STAR:2017kkh} at RHIC, and of 
both charmed hadrons and charmonium bound states in Pb$+$Pb 
collisions~\cite{ALICE:2020pvw,ATLAS:2018xms,CMS:2023mtk,CMS:2016mah,CMS:2017vhp,CMS:2021qqk} 
at the CERN Large Hadron Collider (LHC), indicating that even charm quarks 
diffuse in the QGP 
medium~\cite{Nahrgang:2013xaa,He:2019vgs,Cao:2015hia,Song:2015sfa,Ke:2018jem,Li:2019lex,Plumari:2019hzp}. 
The production of the J/$\psi$ meson, the most abundantly produced 
charmonium bound state, has been extensively studied at both 
RHIC~\cite{PHENIX:2008jgc,PHENIX:2019ihw,PHENIX:2006gsi,PHENIX:2006aub,STAR:2018smh,STAR:2012wnc} 
and the 
LHC~\cite{ALICE:2023gco,ALICE:2023hou,ALICE:2021qlw,CMS:2017dju}. 
However, the elliptic flow of the J/$\psi$ at RHIC energies has 
remained inconclusive, due to the lack of statistical significance.

In relativistic heavy ion collisions, there are several mechanisms that 
may lead to azimuthal anisotropy in J$/\psi$ 
production~\cite{Yao:2018sgn, Yan:2006ve,Greco:2003vf, Zhao:2008vu, 
Liu:2009gx}. The produced J/$\psi$ may be dissolved by the QGP, which creates  
anisotropies in the observed azimuthal distributions due to the 
different path lengths in the medium. If the charm quarks equilibrate 
within the medium, they may coalesce to form J$/\psi$ which may acquire 
significant flow. Additionally, if the J/$\psi$ mesons do not dissolve 
but thermalize inside the medium, they may follow the pressure 
gradients as lighter particles do.

Measurements of the nuclear modification ($R_{AA}$)~\cite{PHENIX:2011img}
of the production of J/$\psi$ can provide insights on these 
competing mechanisms. The $R_{AA}$ is defined as the ratio of the 
production in nucleus-nucleus (AA) collisions to the production in 
proton-proton collisions scaled to account for the number of binary 
nucleon collisions in the AA collisions. When comparing $R_{AA}$ 
measurements at the LHC and RHIC~\cite{ALICE:2016flj}, less suppression 
is observed at the LHC, contrary to expectations, as early theoretical 
calculations of J/$\psi$ predicted that the suppression will become 
stronger as the QGP temperature increases~\cite{Matsui:1986dk}. Also 
unexpected, a greater J/$\psi$ suppression is observed at forward 
rapidity than at midrapidity at RHIC~\cite{PHENIX:2011img}. The 
$R_{AA}$ measurements can be explained~\cite{Zhou:2014kka} by models 
where J/$\psi$ production has a contribution from coalescence of charm 
and anticharm quarks both at RHIC and the LHC. These contributions are more 
significant at the higher center-of-mass energy and at midrapidity, 
where more $c\bar{c}$ pairs are produced. The same models support 
observations of a significant J/$\psi$ $v_2$ at  
the LHC~\cite{ALICE:2020pvw}. The RHIC measurement of J$/\psi$ $v_2$
at midrapidity~\cite{STAR:2012jzy} appears to be consistent with zero, but 
has limited statistical significance. These measurements of 
J/$\psi$ $v_2$ using the dimuon decay channel 
(J/$\psi\rightarrow\mu^++\mu^-$) are the first at forward rapidity at RHIC.

\section{Experimental setup}

Detailed below are the relevant detector subsystems of 
the PHENIX experiment~\cite{PHENIX:2003nhg}, plus a discussion of 
the event and particle selections used in the analysis.

\subsection{PHENIX detectors}
\label{sec:detectors}

\begin{figure}
 \includegraphics[width=1.0\linewidth]{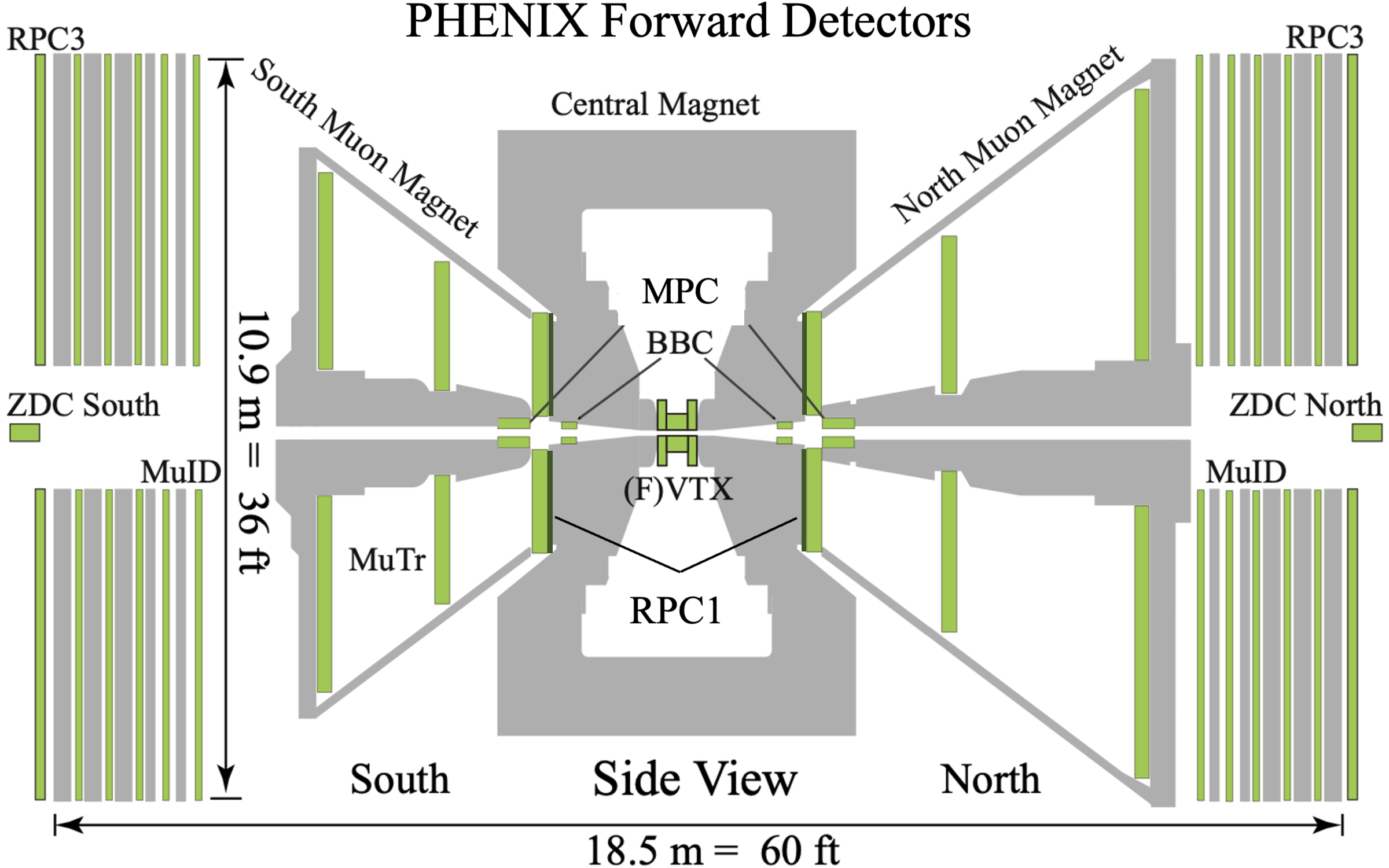}
 \caption{ A side view of the PHENIX muon-arm spectrometers.}
 \label{fig:phenix_detector}
\end{figure}

Figure~\ref{fig:phenix_detector} shows the PHENIX forward detectors 
(north and south arms) used in this analysis, which comprise the muon 
tracker (MuTr), muon identifier (MuID), forward-silicon-vertex detector 
(FVTX) and the beam-beam counter (BBC). The MuTr in each arm covers the 
pseudorapidity ranges of (north) $1.2<\eta <2.4$ and (south) $-2.2<\eta 
<-1.2$ and comprises three octagonal stations of cathode-strip 
chambers. By measuring ionization produced within the gas-filled drift 
chambers, the MuTr reconstructs charged-particle trajectories, and 
subsequently measures their momenta. The MuID comprises five 
alternating layers of steel absorbers and Iarocci tubes. The successive 
hits in the Iarocci tubes form a ``road" that is matched with MuTr 
tracks. The combined information from the MuTr and MuID is used to 
identify muons that form $\mu^{+}\mu^{-}$ pairs to reconstruct the 
mesons J$/\psi$~\cite{PHENIXMUONARMS}.

Two sets of detectors, FVTX and BBC, are used to reconstruct the 
second-order event plane $\Psi_2$ with respect to which the J$/\psi$ 
azimuthal anisotropy is measured.
The FVTX is a four-layer silicon-strip detector placed perpendicular to 
the beam pipe at both sides of the collision point. The strips form 48 
azimuthal segments covering the same rapidity range as the muon 
spectrometers. To avoid short-range correlations that are not 
associated with collective flow, the FVTX in the south arm is used to 
measure the flow of J$/\psi$ in the north arm and vice versa.
The BBCs that detect \v{C}erenkov radiation comprise scintillators and 
photomultiplier tubes surrounding the beam pipe and located between the 
muon arm and the central magnet~\cite{PHENIX_BBC1}. The BBCs cover a 
pseudorapidity region of $3.0<|\eta|<3.9$ and have full azimuthal 
coverage, providing multiplicity and centrality information, 
collision-vertex-position measurements, and reconstruction of the event 
plane.

The PHENIX central-arm tracking detectors (CNT) are drift chambers and 
pad chambers, which also are used for event-plane determination and are 
employed to determine the event-plane resolution as discussed in 
Section~\ref{sec:v2}. Additionally, the central-arm silicon-vertex 
detector (VTX)~\cite{Nouicer:2007rb} aids in accurately locating the 
primary-vertex point along the beam axis.


\subsection{Data samples and track selection}
\label{sec:track_selection}

The integrated luminosity of the PHENIX 2014 and 2016 data sets are 7.5 
nb$^{-1}$ and 7.0 nb$^{-1}$, respectively, with a combined 
${\approx}3.26\times10^{10}$ minimum-bias triggered events. A trigger 
system was employed to select collisions with a $z$-vertex within 
$\pm10$~cm from the center of the detector. Additionally, a dimuon 
trigger based on a combination of MuID hits selected a total of 380M 
dimuon candidates in the 2014 data set, and 307M in the 2016 data set.

Muons are identified by requiring that a detected particle penetrates 
several layers of absorber material in the MuID. By requiring the muon 
candidates to penetrate so far, hadron contribution is greatly reduced, 
although some hadrons may punch through. Particles that make it through 
this threshold, are then matched with the tracks reconstructed by the 
MuTr. Table~\ref{tab:trackvar} summarizes the 
track selections that are implemented to optimize the signal to 
background and were developed through detailed 
Monte-Carlo (MC) studies, 
The tracks in the MuTr must have a minimum of 
six hits and a $\chi^2$/NDF less than 10. The MuID roads must have at 
least three associated hits and MuID roads that have a $\chi^2$/NDF 
larger than 3 are rejected.  The reconstructed muon tracks  
are required to have $p_{T} > 1 $ GeV/$c$, longitudinal momentum $p_{z}> 
3$~GeV/$c$, and track pseudorapidity within $1.2<|\eta|<2.2$. 

Also applied are kinematic cuts to match the particle's trajectory 
in the MuTr track and the MuID road. The distance between the MuTr 
track projection and the first MuID hit and the angle between the MuTr 
track projection and the MuID road linearly depend on the inverse of 
the particle momentum. The muon selection includes a maximum distance 
and angle between the MuTr track projections and MuID roads which are 
scaled by the muon candidate momentum as shown in Table~\ref{tab:trackvar}.

\begin{table}
\caption{\label{tab:trackvar}Summary of muon-track-selection criteria}
\begin{ruledtabular} \begin{tabular}{cc} 
Variable & Cut \\
\hline
$|y|$ & [1.2--2.2] \\ 
$p_T$ & $\geq$1 GeV/$c$ \\ 
$p_z$ & $>$ 3 GeV/$c$\\
MuID penetration layer & 3 or 4 \\ 
Distance difference MuTr \& MuID & $<60$ mm$\cdot$GeV/$c$  \\ 
Angular difference MuTr \& MuID & $<40$ mrad$\cdot$GeV/$c$  \\ 
Track MuTr $\chi^2$ & $<10$ \\ 
Track MuID $\chi^2$ & $<3$ \\ 
\# of MuTr track hits & $>6$ \\ 
\# of MuID hits & $>3$ \\ 
Dimuon vertex refit $\chi^2$ & $<3$
\end{tabular} \end{ruledtabular}
\end{table}

Muon pairs are refitted including the vertex position determined by the 
FVTX, and only pairs providing a refit with $\chi^2<3$ are accepted. 
Additionally, dimuon pairs from decays of J/$\psi$ with $p_T<$5 GeV/$c$ 
are expected to produce tracks with opening angles larger than 45$^o$. 
Therefore an opening angle cut is imposed to those dimuon pairs to 
reduce combinatorial background. Track selections are identical for 
both data sets and after implementation yield 210,389 dimuons in the 
2014, and 140,587 in the 2016 data samples.

\section{Data analysis}

Detailed in this section are discussions of the extraction of the 
J/$\psi$ yield, the measurement of azimuthal anisotropies, and the 
determination of systematic uncertainties

\subsection{J/$\psi$ yield extraction}
\label{sec:yield}

\begin{figure*}
\includegraphics[width=0.99\linewidth]{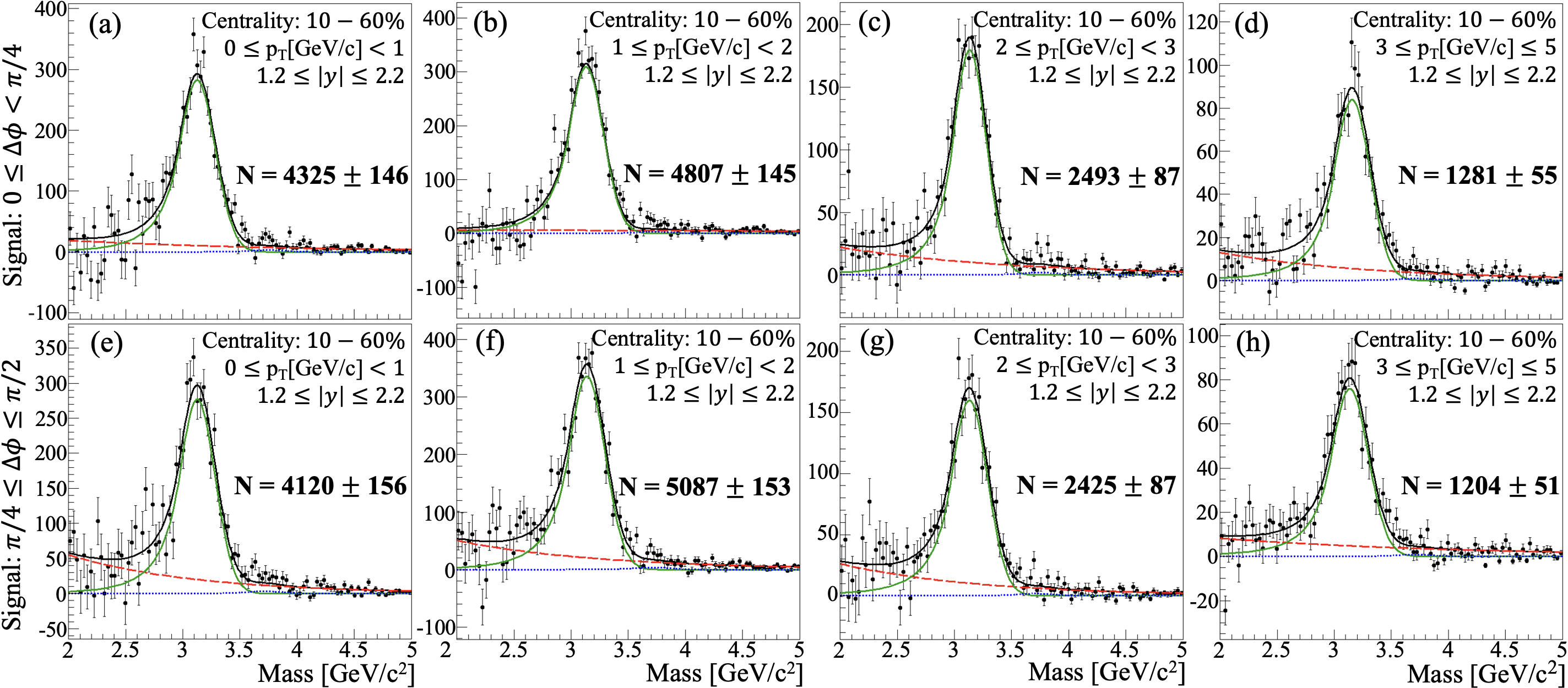}
\caption{Combined 2014 and 2016 data set dimuon invariant-mass 
distributions after mixed-event combinatorial background subtraction. 
The plots are binned by $p_T$ in each column and in $\Delta\phi$ for 
each row.  In decreasing magnitude, the highest [black] curves are the 
total fits, the next highest [green] curves are CB~fits to the 
J/$\psi$ peak, the dashed [red] curves are exponential fits to the 
remaining background after combinatorial-background subtraction, and 
the dotted [blue] curves are CB~fits to the $\psi(2S)$ peak,
}
\label{Fig:F2H}
\end{figure*}

The J$/\psi$ signal is obtained from the invariant-mass distribution of 
muon pairs near the expected J$/\psi$ mass. The large contribution of 
random combinatorial background is estimated using opposite-charge muon 
pairs from different events. 
Mixed-event dimuon pairs are formed only if the two muons have 
differences less than 5\% in centrality, 0.75~cm in $z$-vertex, and 
$\pi/20$~rad in event-plane angle.  Otherwise, a mixed-event muon 
pair is not formed.
The sample of mixed-event dimuons is about four times larger than 
dimuons from the same event. A normalization factor must be applied for 
the mixed-event sample, which can be obtained by using the ratio of 
like-sign pairs from the same event to like-sign pairs from 
mixed-events. The signal is then obtained by the subtraction of the 
normalized background from the foreground, which results in increased 
statistical precision when compared to the like-sign subtraction, and 
is of the form:

\begin{equation}
S = {\rm FG}-N\cdot {\rm BG}^{\rm mix},
\end{equation}

\noindent where $S$ is the signal, FG is the foreground distribution 
using opposite-sign muons from the same event, BG$^{\rm mix}$ is the 
background distribution created by mixing events, and $N$ is the 
normalization factor that is obtained using:

\begin{equation}
N^{\rm mix} = \frac{\sqrt{N^{++}_{\rm same}\cdot N^{--}_{\rm same}}}
{\sqrt{N^{++}_{\rm mix}\cdot N^{--}_{\rm mix}}},
\end{equation}

\noindent where $N^{++}$ and $N^{--}$ denote the number of like-sign 
muon pairs, while ``same" and ``mix" denote whether the pair came from 
the same event or from event-mixing.  $N^{\rm mix}$ is the resulting 
distribution, which has a mass dependence and was fit with a polynomial 
of degree two. The minimum of the function was taken to be $N$ used to 
scale the combinatorial background, $N={\rm min}(N_{\rm mix})$. The 
uncertainty associated with the fit was determined to be 5\% and is 
propagated as a systematic uncertainty. In the fitting of the mass 
distributions, the shape of the J/$\psi$ signal is assumed to be a 
crystal-ball (CB) function~\cite{Gaiser:1982yw}, and the J/$\psi$ count is 
obtained by the integral of the CB function in the fit (see 
Fig.~\ref{Fig:F2H}).

\subsection{Azimuthal anisotropy measurement}
\label{sec:v2}

In a collision of two nuclei, the reaction plane is defined as the 
plane formed by the impact-parameter vector and the beam axis. In 
practice, because this reaction plane is not directly observable, the 
event-plane method is employed that uses the anisotropic flow itself to 
estimate the reaction-plane angle~\cite{Poskanzer:1998yz}. The $x$ and 
$y$ components of the event-flow vector $Q_n$ and the event-plane 
angle $\Psi_n$ are defined for each $n^{th}$ harmonic by the following:

\begin{equation}
 Q_x=\sum^N_{i=1}\omega_i\cos(n\phi_i),
\end{equation} 
 
\begin{equation}
 Q_y=\sum^N_{i=1}\omega_i\sin(n\phi_i),
\end{equation}

\begin{equation}
 \Psi_n = \frac{1}{n}\tan^{-1}\left(\frac{Q_y}{Q_x}\right).
\end{equation}

\noindent where the Q-vectors are measured by the sum of the vector 
components of the azimuthal angles ($\phi$) of each particle $i$ in an 
event, and $\omega_i$ are weights that are set to 1. The observed 
$v^{\rm obs}_n$ with respect to the event plane is defined by:

\begin{equation}
 v_n^{\rm obs}(p_T,y)=\langle \cos[n(\phi_i-\Psi_n)]\rangle.
\end{equation}

\noindent where $\langle \rangle$ denotes an average over all particles 
and all events. The second-order event-plane angle $\Psi_2$ used in 
measuring $v_2$ is determined using the FVTX (south and north) detectors. 
The finite resolution in the detector and the finite number of 
particles used in event-plane determination require a resolution 
correction to $v_2^{\rm obs}$:

\begin{equation}
 v_2=\frac{v_2^{\rm obs}(p_T,y)}{{\rm Res(\Psi_2)}}.
\end{equation}

The associated event-plane resolutions (see Table~\ref{tab:eprfvtx}) 
are calculated using the standard three-subevent 
method~\cite{Poskanzer:1998yz}, where the particles used in the 
event-plane determination are sorted in three regions A, B, C 
separated in pseudorapidity. Independent measurements are correlated 
using the FVTX, BBC, and CNT, for example:

\begin{equation}
 \rm Res(\Psi^{A}_2)=\sqrt{\frac{\langle \cos(2[\Psi^A_2-\Psi^B_2])\rangle \langle \cos(2[\Psi^A_2-\Psi^C_2])\rangle}{\langle \cos(2[\Psi^B_2-\Psi^C_2])\rangle}}
\end{equation}

\noindent where the $\Psi^A_2$ is the second-harmonic event plane 
measured with the FVTX (either south or north), $\Psi^B_2$ is from the 
BBC (opposite arm north or south), and $\Psi^C_2$ is from the CNT.

To measure $v_2^{\rm obs}$ the foreground and background dimuons are 
split into two bins of $\Delta\phi=(\phi_i-\Psi_2)$, where dimuons from 
$\Delta\phi=[-\pi/4,\pi/4]$ are considered to be ``in-plane" with the 
event plane, and dimuons from $\Delta\phi=[-\pi/2,-\pi/4]$ and 
$[\pi/4,\pi/2]$ are ``out-of-plane."  The J$/\psi$ signals are 
reconstructed for both sets to obtain the yield (See 
Table~\ref{tab:jpsicount}), and then the J/$\psi$ yields are fit as a 
function of $\Delta\phi$ with the following equation:

\begin{equation}
 f(\Delta\phi)=N_{0}(1+2v_2^{\rm obs}\cos(2\Delta\phi)),
  \end{equation}

\noindent where $N_{0}$ is a normalization factor, and $v_2^{\rm obs}$ 
is the free-elliptic-flow parameter. After extracting $v_2^{\rm obs}$ 
for each data set, the event-plane resolution is corrected. A weighted 
average between the 2014 and 2016 datasets is obtained using the 
statistical uncertainties of the two measurements as weights. Repeating 
this for each measurement evaluates the systematic uncertainties.  

\begin{table}
\caption{\label{tab:eprfvtx}Event-plane resolutions using the FVTX as 
the primary event-plane detector.}
\begin{ruledtabular} \begin{tabular}{ccc}
 Centrality &  $Res(\Psi^{\rm FVTX}_2)$ & $Res(\Psi^{\rm BBC}_2)$ \\
 \hline
 0\%--10\%  & 0.3035 & 0.1833 \\
 10\%--20\% & 0.5359 & 0.2637 \\
 20\%--30\% & 0.5701 & 0.2803 \\
 30\%--40\% & 0.5391 & 0.2582 \\
 40\%--50\% & 0.4639 & 0.2108 \\
 50\%--60\% & 0.3620 & 0.1546 \\
\end{tabular} \end{ruledtabular}
\end{table}

\begin{table}
\caption{\label{tab:jpsicount}
The J/$\psi$ yields recorded in 2014 and 2016 within the in-plane (in) and 
out-of-plane (out) $\Delta\phi$ bins.}
\begin{ruledtabular} \begin{tabular}{cccc} 
Year &  $p_T$ bin (GeV/$c$) &  $N_{\rm in}$ & $N_{\rm out}$ \\
\hline
2014
& 0.0--1.0 & $2727\pm100$ & $2454\pm107$ \\
& 1.0--2.0 & $2863\pm145$ & $3136\pm110$ \\
& 2.0--3.0 & $1636\pm65 $ & $1456\pm63$  \\
& 3.0--5.0 & $ 791\pm42 $ & $ 780\pm38$  \\
 \\
& 0.0--0.5 & $ 832\pm65 $ & $ 770\pm72$  \\
& 0.5--1.0 & $1872\pm82 $ & $1648\pm87$  \\
& 1.0--5.0 & $5415\pm146$ & $5373\pm127$ \\
 \\
& 0.0--5.0 & $8189\pm160$ & $7845\pm164$  \\
 \\
2016
& 0.0--1.0 & $1498\pm125$ & $1685\pm88$  \\
& 1.0--2.0 & $1930\pm79$  & $1985\pm95$  \\
& 2.0--3.0 & $ 937\pm54$  & $ 970\pm51$  \\
& 3.0--5.0 & $ 493\pm32$  & $ 438\pm31$  \\
 \\ 
& 0.0--0.5 & $ 470\pm50$  & $ 643\pm50$  \\
& 0.5--1.0 & $1025\pm81$  & $1026\pm75$  \\
& 1.0--5.0 & $3399\pm100$ & $3398\pm109$ \\
 \\
& 0.0--5.0 & $4965\pm131$ & $5092\pm136$ \\
\end{tabular} \end{ruledtabular}
\end{table}

\subsection{Systematic uncertainties}
\label{sec:syst}

Below are detailed descriptions of the sources of systematic uncertainty.

\subsubsection{Background contributions} 

The like-sign method is an alternative to estimate the combinatorial 
background.  The background is created using dimuon pairs of the same 
sign ($\mu^+\mu^+$ and $\mu^-\mu^-$) that come from the same event:

\begin{equation}
BG = 2R\sqrt{N^{++}N^{--}},
\end{equation}

\noindent where $N^{++}$ and $N^{--}$ are the like-sign dimuon counts 
per mass bin, and $R$ is a normalization factor that is obtained by 
taking a ratio of the foreground over background in a region where a 
signal ($M_{\mu\mu}=$[1.5--2.5]~GeV/$c^2$) is not expected.  The 
contribution to the final systematic uncertainty is small, indicating 
good agreement between the two background-estimation methods.

Because the measurement using mixed-event BG estimation has smaller 
uncertainties in the final $v_2$ measurement, the default is chosen to
be mixed-event subtraction.  The uncertainty of the mixed-event background 
normalization was determined to be $\pm5\%$. This is propagated to the 
final systematic uncertainty by varying the normalization by $\pm5\%$, 
reconstructing the J/$\psi$ for these two sets of mass distributions, 
and calculating the resulting $v_2$ for each set.

\subsubsection{Track selections} 

To evaluate the uncertainty from the track selections listed in 
Table~\ref{tab:trackvar}, the selections were varied to make them tighter 
or looser by 10\% and the analysis was performed to obtain $v_2$ as in 
the default measurement. 
Table~\ref{tab:syscontribution} shows the maximum absolute difference 
in $v_2$ measured in each $p_T$ and centrality interval.

\subsubsection{Detector effects} 

The J$/\psi$ efficiency is not uniform across the measured $p_T$ and 
rapidity ranges and if the $v_2$ signal is nonzero or has a $p_T$ and 
rapidity dependence within the measured bins, the $v_2$ measurements 
will be distorted.  These distortions were evaluated in simulations and 
propagated as systematic uncertainties. A sample of J/$\psi$ decays 
were simulated and reconstructed with a full {\sc 
geant4}~\cite{GEANT4:2002zbu} detector simulation. The J/$\psi$ sample 
was weighted according to the yield measured by 
PHENIX~\cite{PHENIX:2006gsi}. The event-plane angle was set to 0 and 
the J/$\psi$ azimuthal angle distributions was modulated according to:

\begin{equation}
 f(\phi) =1+2 v_2^{\rm gen}(p_T, \eta, {\rm centrality})\cos(2\phi), 
\end{equation}

\noindent where $v_2^{\rm gen}$ are the $v_2$ measured by the 
CMS~\cite{CMS:2017xgk,CMS:2017vhp} and ALICE~\cite{ALICE:2020pvw} 
detectors at the LHC.  The LHC measurements are used to estimate an 
upper limit for our J/$\psi$ $v_{2}$ signal. The difference between the 
measured $v_2$ of reconstructed J/$\psi$ from this sample and 
$v_2^{\rm gen}$ is no larger than 0.002, which is included in the systematic 
uncertainty.

\subsubsection{Fourier fitting} 

The uncertainties associated with the coarse binning in the $\Delta 
\phi$ distributions are used to obtain the $v_{2}$ values. As a 
systematic check, the J/$\psi$ yields are split into three 
$|\Delta\phi|$ bins ([0--$\pi/6$], [$\pi/6$--$\pi/3$], and 
[$\pi/3$--$\pi/2$]) and are fit to the $|\Delta\phi|$ distributions 
with Eq. 10. After extracting $v_2^{\rm obs}$ for each data set, the 
event-plane resolutions are corrected.  The absolute difference between 
the $v_2$ values obtained with two or three $\Delta\phi$ bins are taken 
as a systematic uncertainty and listed in 
Table~\ref{tab:syscontribution} as $\Delta \phi$ Bin.

\subsubsection{Event-plane detector} 

In the standard analysis the J$/\psi$ mesons detected in the south 
(north) muon spectrometer are correlated with the event plane 
determined in the FTVX from the opposite arm, north (south). The 
selection imposes a pseudorapidity gap of $2.4<|\Delta\eta|<4.4$ 
between the particles used in the flow analysis and those used to 
determine the event-plane angle. As a systematic check, the event-plane 
angle determined in the north (south) FVTX is replaced with the angle 
measured in BBC north (south). The resulting $v_2^{\rm obs}$ 
measurement is corrected with the BBC event-plane resolution. The 
absolute difference between the two $v_2$ measurements is propagated as 
a systematic uncertainty.

\subsubsection{Summary of systematic uncertainties} 

\begin{table*}
\caption{Summary of systematic uncertainty contributions for 
centralities 10\%--60\%, 10\%--40\%, and 0\%--50\%. See text for
descriptions of each contribution.}
\label{tab:syscontribution}
\begin{ruledtabular}  \begin{tabular}{cccccccccc}
Centrality & ${\rm p}_{T}$(GeV/$c$) & LSS & $\textbf{N}^{\rm mix}$ & Track Selection & EFF & $\Delta\phi$ Bin & EP & Total \\
 \hline
10\%--60\% 
& 0.0--1.0& $0.0133$ & $0.0107$ & $0.0025$ & $0.0020$ & $0.0022$ & $0.0038$ & $0.018$\\
& 1.0--2.0& $0.0051$ & $0.0172$ & $0.0091$ & $0.0020$ & $0.0114$ & $0.0015$ & $0.023$\\
& 2.0--3.0& $0.0015$ & $0.0006$ & $0.0019$ & $0.0020$ & $0.0046$ & $0.0005$ & $0.005$\\
& 3.0--5.0& $0.0042$ & $0.0003$ & $0.0022$ & $0.0020$ & $0.0081$ & $0.0102$ & $0.014$\\
 \\
& 0.0--0.5& $0.0004$ & $0.0003$ & $0.0005$ & $0.0020$ & $0.0004$ & $0.0001$ & $0.001$\\
& 0.5--1.0& $0.0061$ & $0.0018$ & $0.0012$ & $0.0020$ & $0.0012$ & $0.0049$ & $0.008$\\
& 1.0--5.0& $0.0019$ & $0.0078$ & $0.0040$ & $0.0020$ & $0.0120$ & $0.0011$ & $0.015$\\
 \\
& 0.0--5.0& $0.0059$ & $0.0085$ & $0.0045$ & $0.0020$ & $0.0082$ & $0.0013$ & $0.014$\\
 \\
10\%--40\% 
& 0.0--2.0 & $0.0214$& $0.0145$& $0.0058$ & $0.0020$ & $0.0105$ & $0.0052$ & $0.029$\\
& 2.0--5.0 & $0.0265$& $0.0003$& $0.0128$ & $0.0020$ & $0.0055$ & $0.0002$ & $0.025$\\
& 5.0--10.0 & $0.0178$& $0.0006$& $0.0263$ & $0.0020$ & $0.0056$ & $0.0677$ & $0.075$\\
 \\
0\%--50\% 
& 0.0--2.0 & $0.0116$& $0.0149$ & $0.0214$ &$0.0020$& $0.0015$ & $0.0144$ &$0.032$\\
& 2.0--5.0 & $0.0084$& $0.0000$ & $0.0032$ &$0.0020$& $0.0006$ & $0.0064$ &$0.011$\\
& 5.0--10.0 & $0.0142$& $0.0004$ & $0.0216$ &$0.0020$& $0.0079$ & $0.0591$ &$0.065$\\
\end{tabular} \end{ruledtabular} 
\end{table*}

Table~\ref{tab:syscontribution} presents  
the contributions of each of the systematic uncertainties 
for J$/\psi$ $v_2$ as a function of $p_T$ in 
the PHENIX muon arms ($1.2<|\eta|<2.2$) for the 
centrality selections 10\%--60\%, 0\%--50\%, and 10\%--40\%). 
The descriptions for each contribution are:

\begin{description}

\item[LSS] 
The contribution from using like-sign background subtraction

\item[N$^{\rm mix}$] 
The contribution from altering the mixed-event normalization by the 
associated uncertainty of determining the normalization factor 

\item[Track Selection] 
The contribution from altering the track quality cuts from 
Table~\ref{tab:trackvar} by 10\% of their default values

\item[EFF] 
The contribution from nonuniform detector acceptance

\item[$\Delta\phi$ Bin]  
The contribution from the selection of $\Delta\phi$ binning in the 
Fourier fits

\item[EP] 
The contribution from the event-plane 
detector selection

\item[Total] 
The total systematic uncertainty of the contributions, which are 
assumed to be independent, for each $p_T$ bin and added in quadrature

\end{description}

\section{Results and discussion}
\label{sec:results}

\begin{figure}
\begin{minipage}{0.995\linewidth}
\vspace{0.2cm}
 \includegraphics[width=1.0\linewidth]{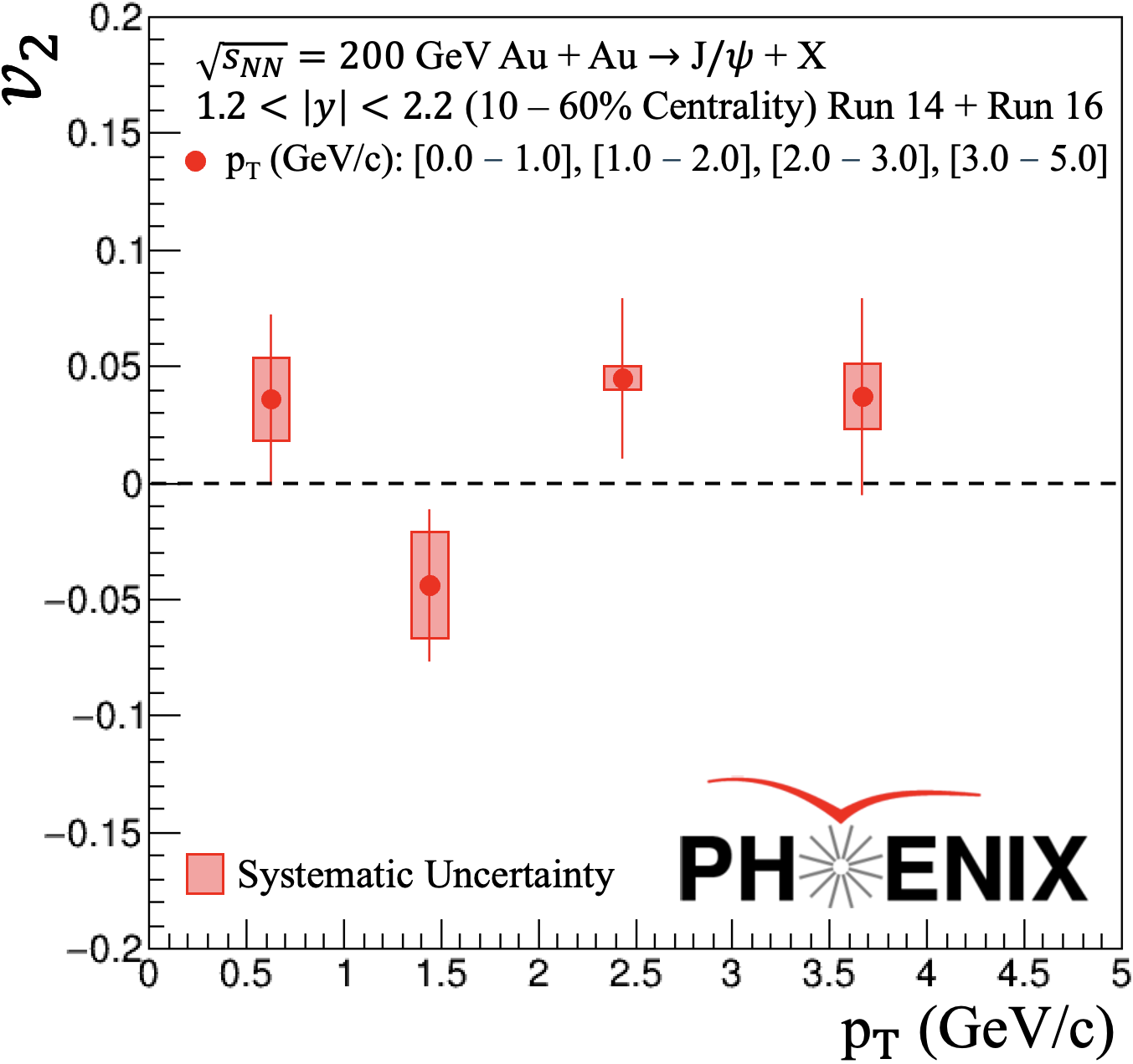}
 \caption{Measurement of J/$\psi$ $v_2$ at forward rapidity 
($1.2<|\eta|<2.2$) for centrality 10\%--60\% in Au$+$Au collisions at 
$\sqrt{s_{_{NN}}}=200$~GeV. The measurements are performed for the 
$p_T$ selections [0.0--1.0], [1.0--2.0], [2.0--3.0], and 
[3.0--5.0] GeV/$c$. The data points are plotted at the mean $p_T$ 
for each bin.
}
 \label{fig:default}
\end{minipage}
\begin{minipage}{0.995\linewidth}
\vspace{0.5cm}
 \includegraphics[width=1.0\linewidth]{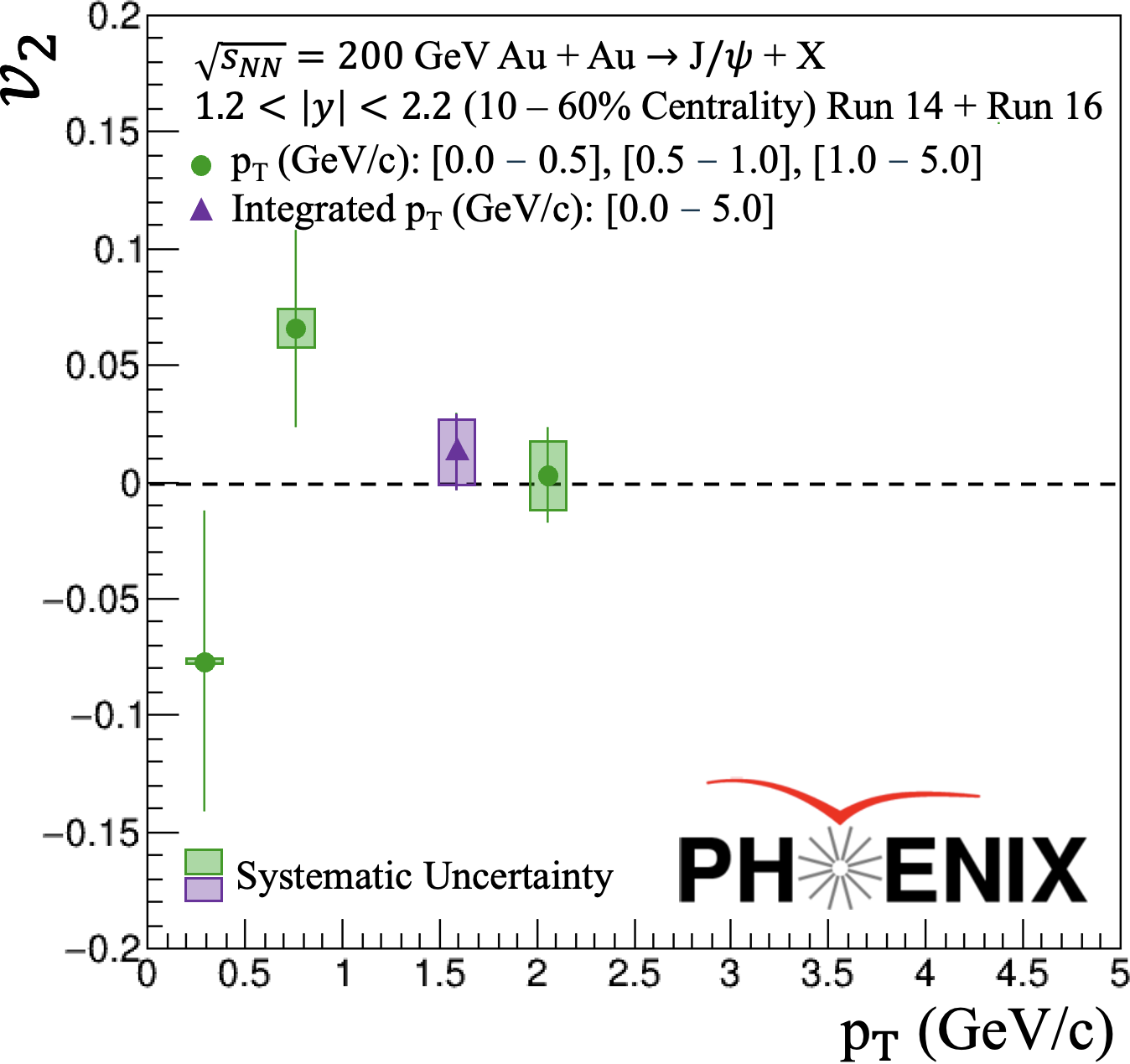}
 \caption{Measurement of $v_2$ of J/$\psi$ at forward rapidity 
($1.2<|\eta|<2.2$) for centrality 10\%--60\% in Au$+$Au 
collisions at $\sqrt{s_{_{NN}}}=200$~GeV. 
The solid circle [green] data points are measurements for the 
$p_T$ selections [0.0--0.5], [0.5--1.0], and 
[1.0--5.0]~GeV/$c$.  The triangle [dark blue] data point at
$p_T=1.6$~GeV/$c$ 
is for the integrated $p_T$ selection [0.0--5.0]~GeV/$c$.
All data points are plotted at the mean $p_T$ for each bin.
}
\label{fig:special}
\end{minipage}
\end{figure}

In the first centrality range (10\%--60\%) of 
Table~\ref{tab:syscontribution}, the $v_2$ was measured for three 
configurations of $p_T$ (see Fig.~\ref{fig:default} and 
Fig.~\ref{fig:special}): (1) standard binning ([0.0--1.0], [1.0--2.0], 
[2.0--3.0], and [3.0--5.0]~GeV/$c$), (2) special binning ([0.0--0.5], 
[0.5--1.0], and [1.0--5.0]~GeV/$c$), and (3) with one inclusive $p_T$ 
bin ([0.0--5.0] GeV/$c$). In each case, the measured J$/\psi$ $v_2$ 
is consistent with zero within uncertainties, and the numerical values 
are shown in Table~\ref{tab:final}.

\begin{figure}
 \includegraphics[width=1.0\linewidth]{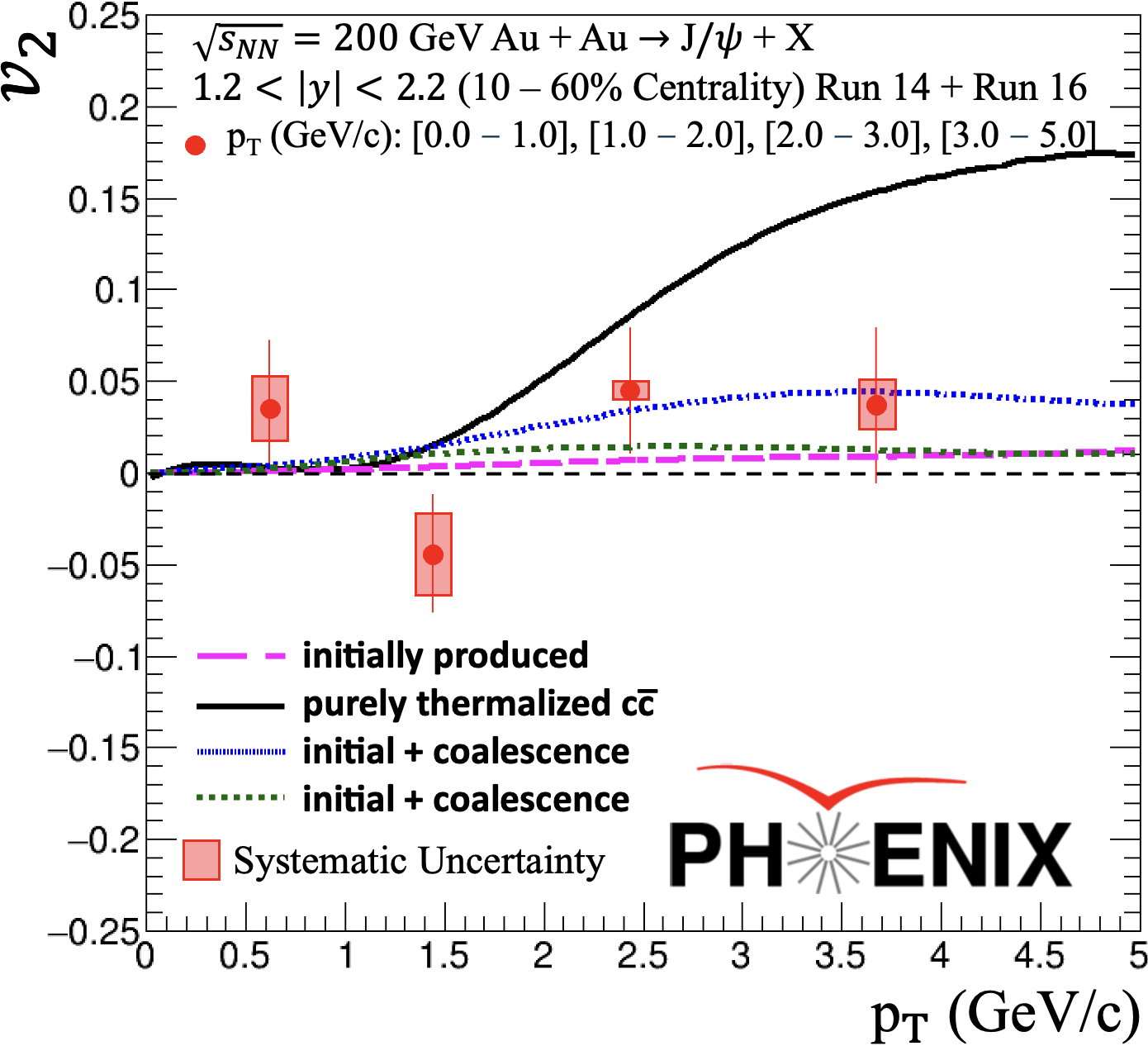}
 \caption{Forward-rapidity J/$\psi$ $v_2$  as a function of $p_T$ for 
centrality 10\%--60\% in Au$+$Au collisions at 
$\sqrt{s_{_{NN}}}=200$~GeV compared to various theoretical
models~\protect\cite{Yan:2006ve,Greco:2003vf,Zhao:2008vu,Liu:2009gx}. 
The mean $p_T$ for each bin is used for each measurement.}
\label{fig:theory}
\end{figure}

\begin{table}
\caption{\label{tab:final} Values of J/$\psi$ $v_2$ measured in the 
10\%--60\%, 10\%--40\%, and 0\%--50\% centrality ranges with 
statistical and systematic uncertainties.  The 10\%--40\% and 0\%--50\% 
centrality ranges are used to facilitate comparisons with previous 
measurements by STAR~\cite{STAR:2012jzy} 
(See Fig.\protect\ref{fig:STAR}) and ALICE~\cite{ALICE:2020pvw} 
(See Fig.\protect\ref{fig:ALICE}), respectively.
}
\begin{ruledtabular} \begin{tabular}{ccc}
$\langle p_T\rangle$ [GeV/$c$] & J/$\psi$ $v_{2}$ & Measurement\\
\hline
10\%--60\%
& 0.62 & $0.036\pm0.036^{\rm stat}\pm0.018^{\rm sys}$  \\
& 1.44 & $-0.044\pm0.032^{\rm stat}\pm0.023^{\rm sys}$ \\
& 2.43 & $0.045\pm0.034^{\rm stat}\pm0.005^{\rm sys}$  \\
& 3.67 & $0.037\pm0.042^{\rm stat}\pm0.014^{\rm sys}$  \\
 \\
& 0.30 & $-0.077\pm0.064^{\rm stat}\pm0.001^{\rm sys}$ \\
& 0.76 & $0.066\pm0.042^{\rm stat}\pm 0.008^{\rm sys}$ \\
& 2.06 & $0.003\pm0.020^{\rm stat}\pm 0.015^{\rm sys}$ \\
 \\
& 1.59 & $0.013\pm0.016^{\rm stat}\pm0.014^{\rm sys}$ \\
10\%--40\% 
& 1.04 & $ 0.001\pm0.026^{\rm stat}\pm0.029^{\rm sys}$ \\
& 3.00 & $ 0.051\pm0.035^{\rm stat}\pm0.025^{\rm sys}$ \\
& 6.45 & $-0.201\pm0.133^{\rm stat}\pm0.075^{\rm sys}$ \\
 \\
0\%--50\%
& 1.04 & $-0.028\pm0.039^{\rm stat}\pm0.032^{\rm sys}$ \\
& 3.00 & $ 0.043\pm0.036^{\rm stat}\pm0.011^{\rm sys}$ \\
& 6.41 & $-0.048\pm0.151^{\rm stat}\pm0.065^{\rm sys}$ \\
\end{tabular} \end{ruledtabular} 
\end{table}

\begin{figure}
\begin{minipage}{0.995\linewidth}
\vspace{0.2cm}
 \includegraphics[width=1.0\linewidth]{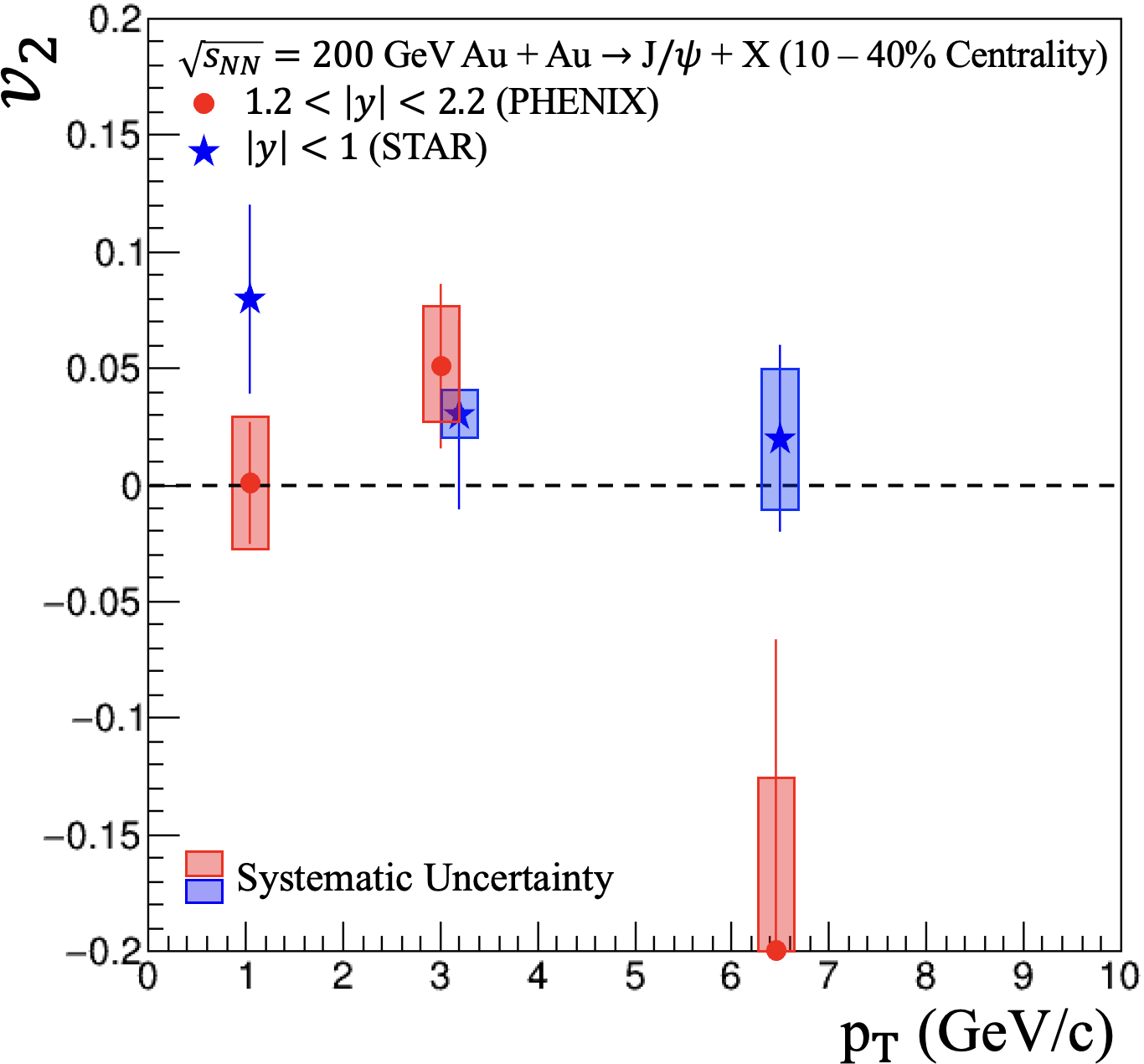}
 \caption{Comparison of measurements between PHENIX (round [red] data 
points) at forward rapidity and STAR~\cite{STAR:2012jzy} (star-symbol 
[blue] data points) at midrapidity for J/$\psi$ $v_2$ as a function of 
$p_T$ for centrality 10\%--40\% in Au$+$Au collisions at 
$\sqrt{s_{_{NN}}}=200$~GeV.  The $p_T$ bins for both measurements are 
[0.0--2.0], [2.0--5.0], and [5.0--10.0]~GeV/$c$, and all data are 
plotted at the mean $p_T$ for each bin.
}
 \label{fig:STAR}
\end{minipage}
\begin{minipage}{0.995\linewidth}
\vspace{0.5cm}
 \includegraphics[width=1.0\linewidth]{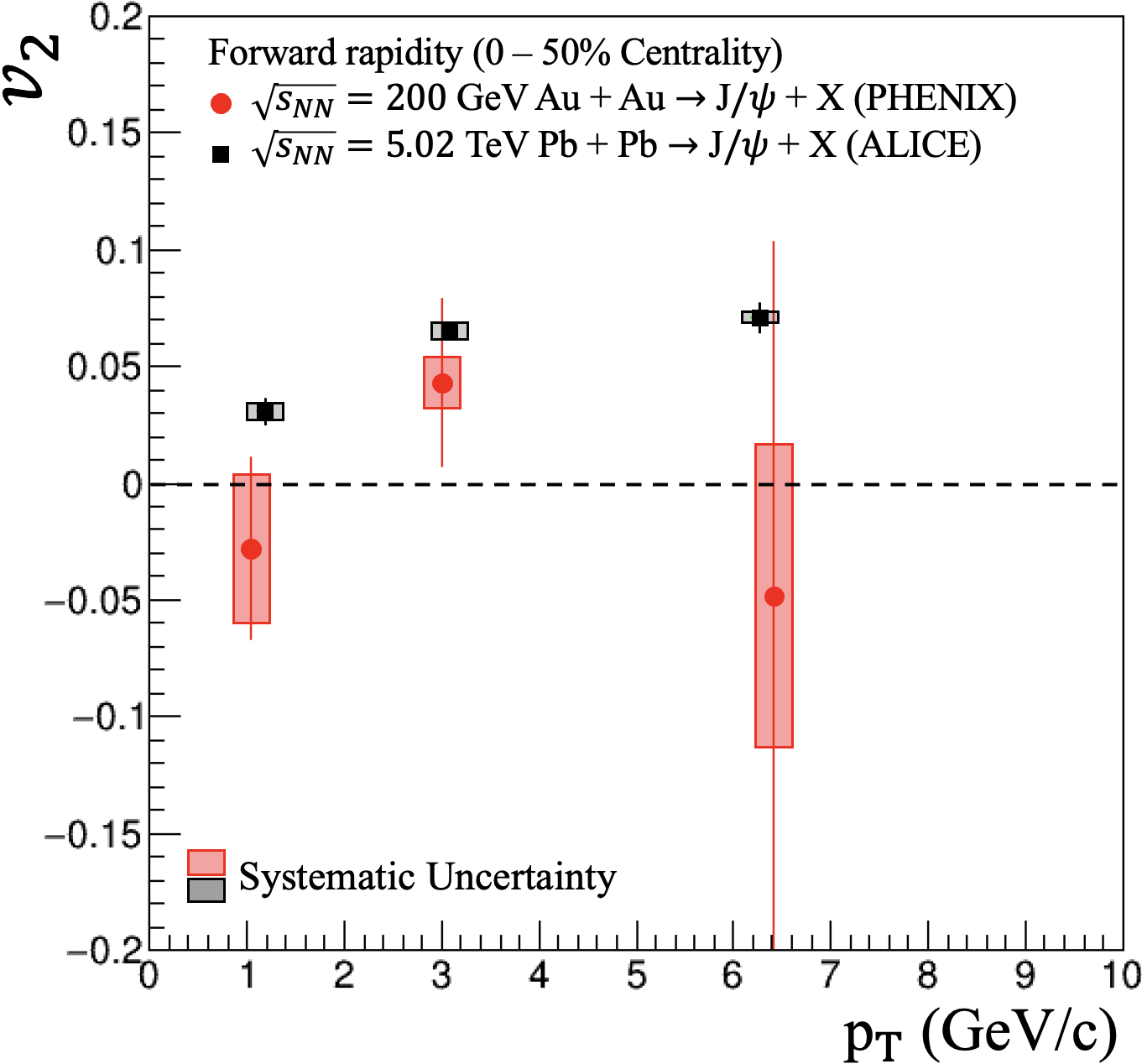}
 \caption{The J/$\psi$ $v_2$ as a function of $p_T$ for centrality 
0\%--50\% at forward rapidity. The circle [red] data points are for 
PHENIX Au$+$Au collisions at $\sqrt{s_{_{NN}}}=200$~GeV for 
$1.2<|y|<2.2$ and the square [black] data points are for ALICE Pb$+$Pb 
collisions at $\sqrt{s_{_{NN}}}=5.02$~TeV for 
$2.4<|y|<4$~\cite{ALICE:2020pvw}. The $p_T$ bins for both measurements 
are [0.0--2.0], [2.0--5.0], and [5.0--10.0]~GeV/$c$. The data points 
are plotted at the mean $p_T$ for each bin.
}
 \label{fig:ALICE}
\end{minipage}
\end{figure}

Figure~\ref{fig:theory} compares the PHENIX measurement of J/$\psi$ 
elliptic flow at forward rapidity to several theoretical 
models~\cite{Yan:2006ve,Greco:2003vf,Zhao:2008vu,Liu:2009gx} that were 
previously compared to the STAR measurement of J/$\psi$ $v_{2}$ at 
midrapidity~\cite{STAR:2012jzy}. The J$/\psi$ production mechanisms 
vary for each of these models, with the ``initially produced" and 
``coalescence from thermalized $c\bar{c}$" models being at the opposite 
ends of the spectrum. In the initially-produced 
model~\cite{Yan:2006ve}, J$/\psi$ mesons are produced in the initial 
hard scattering without the consideration of regeneration in the QGP. 
In this case azimuthal anisotropy can be generated through 
path-length-dependent dissociation of J$/\psi$ in the medium. This 
effect appears to be small.

As an upper limit of the J$/\psi$ flow, the authors of 
ref.~\cite{Greco:2003vf} consider the production of J$/\psi$ from 
coalescence of thermalized $c$ and $\bar{c}$ at hadronization, where 
the quark elliptic flow is fully developed. 
Alternatively~\cite{Yan:2006ve}, the $c$ and $\bar{c}$ coalescence can 
be considered throughout the QGP evolution which will allow for 
charmonium creation at earlier stages when the quark flow is smaller. 

In a realistic scenario, both initial production and coalescence 
must be included. 
Two-component models that consider the interplay of both mechanisms 
were employed in refs.~\cite{Zhao:2008vu,Liu:2009gx}. In both cases 
coalescence plays a significant role in J$/\psi$ production at 
low-$p_T$, and the higher $p_T$ (${\approx}2$~GeV/$c$) is dominated by 
perturbative quantum chromodynamics. In contrast, the flow 
of the charm quarks is boosted to higher $p_T$ than that of the light 
quarks due to radial flow leading to overall near-zero elliptic flow 
for J$/\psi$ with $p_{T}<1$~GeV/$c$. Combining the two mechanisms leads 
to nonzero $v_2$ of J$/\psi$ at higher $p_T$ peaking around 
$p_T\approx$3~GeV/$c$, but the magnitude is small. In the scenario 
where $c\bar{c}$ coalescence happens at 
hadronization~\cite{Zhao:2008vu} the $v_2$ of J$/\psi$ reaches up to 
3\%, but is smaller in the case of continuous 
coalescence~\cite{Liu:2009gx}. Our data exclude a scenario in which  
J$/\psi$ are produced entirely from coalescence of thermalized charm 
quarks. However, the models cannot be distinguished from each other 
when including only primordial J$/\psi$ or production through 
coalescence.


To facilitate direct comparisons to previous measurements of J/$\psi$ 
elliptic flow at RHIC and the LHC, measurements are performed for 
other centrality and $p_T$ selections ([0.0--2.0], [2.0--5.0], and 
[5.0--10.0]~GeV/$c$).  Figure~\ref{fig:STAR} compares, in the 
centrality range of 10\%--40\%, this PHENIX measurement at forward 
rapidity with the STAR midrapidity result~\cite{STAR:2012jzy}. A 
significant rapidity dependence or significant elliptic flow of 
J/$\psi$ is not observed in Au$+$Au collisions at 
$\sqrt{s_{_{NN}}}=200$ GeV. Figure~\ref{fig:ALICE} compares the PHENIX 
and ALICE results~\cite{ALICE:2020pvw}, where both measurements are 
performed at forward rapidity and in the centrality range of 
0\%--50\%. The PHENIX measurement is systematically lower than the 
$v_{2}$ values observed by ALICE.  Note that the PHENIX measurements 
are consistent with zero elliptic flow, while there is a clear signal 
in the LHC measurements. The numeric values of $v_2$ used in the 
comparisons with previous measurements are shown in 
Table~\ref{tab:final}.

\section{Summary and conclusions}
\label{sec:summary}

In summary, elliptic-flow measurements of inclusive J$/\psi$ 
production are presented as a function of $p_T$ at forward rapidity 
($1.2\leq|y|\leq2.2$) in Au$+$Au collisions at 
$\sqrt{s_{_{NN}}}=200$~GeV. The J$/\psi$ mesons were reconstructed in 
$\mu^{+}\mu^{-}$ decay channel using the PHENIX muon-arm spectrometers. 
The $v_2$ measurements were performed with the event-plane method. The 
event plane was determined using the north and south FVTX detectors in 
the opposite muon arm (south or north) where the J$/\psi$ was measured. 
The PHENIX Au$+$Au data sets collected in 2014 and 2016 sampled 
integrated luminosity of 14.5 $nb^{-1}$. The measurements were 
performed for several centrality selections (10\%--60\%, 10\%--40\%, 
and 0\%--50\%) and compared both to theoretical models and to earlier 
measurements for midrapidity at RHIC and for forward rapidity at the 
LHC. 

In conclusion, The PHENIX measurements at forward rapidity are 
consistent with zero, which is similar to the midrapidity measurement 
from STAR. This is distinct from the ALICE results at the LHC, which 
support significant J$/\psi$ production through charm-quark 
coalescence. The PHENIX data exclude a large contribution at RHIC 
energies to J$/\psi$ production and flow at forward rapidity from 
coalescence of thermalized $c\bar{c}$ quarks.


\begin{acknowledgments}

We thank the staff of the Collider-Accelerator and Physics
Departments at Brookhaven National Laboratory and the staff of
the other PHENIX participating institutions for their vital
contributions.  
We acknowledge support from the Office of Nuclear Physics in the
Office of Science of the Department of Energy,
the National Science Foundation,
Abilene Christian University Research Council,
Research Foundation of SUNY, and
Dean of the College of Arts and Sciences, Vanderbilt University
(U.S.A),
Ministry of Education, Culture, Sports, Science, and Technology
and the Japan Society for the Promotion of Science (Japan),
Natural Science Foundation of China (People's Republic of China),
Croatian Science Foundation and
Ministry of Science and Education (Croatia),
Ministry of Education, Youth and Sports (Czech Republic),
Centre National de la Recherche Scientifique, Commissariat
{\`a} l'{\'E}nergie Atomique, and Institut National de Physique
Nucl{\'e}aire et de Physique des Particules (France),
J. Bolyai Research Scholarship, EFOP, HUN-REN ATOMKI, NKFIH,
and OTKA (Hungary), 
Department of Atomic Energy and Department of Science and Technology (India),
Israel Science Foundation (Israel),
Basic Science Research and SRC(CENuM) Programs through NRF
funded by the Ministry of Education and the Ministry of
Science and ICT (Korea).
Ministry of Education and Science, Russian Academy of Sciences,
Federal Agency of Atomic Energy (Russia),
VR and Wallenberg Foundation (Sweden),
University of Zambia, the Government of the Republic of Zambia (Zambia),
the U.S. Civilian Research and Development Foundation for the
Independent States of the Former Soviet Union,
the Hungarian American Enterprise Scholarship Fund,
the US-Hungarian Fulbright Foundation,
and the US-Israel Binational Science Foundation.

\end{acknowledgments}



%
 
\end{document}